\documentclass[pre,twocolumn]{revtex4} 

\usepackage{graphicx}
\usepackage{amsmath}
\usepackage{amsfonts} 
\usepackage{mathrsfs} 
\usepackage{color}
\usepackage{natbib,hyperref,url}
\usepackage{bm}

\def\<{\langle}
\def\>{\rangle}

\def\beq{\begin{equation}}
\def\eeq{\end{equation}}
\def\barray{\begin{eqnarray}}
\def\earray{\end{eqnarray}}

\def\im{{\rm Im}\,}

\def\bc{\beta_{\rm t}}
\def\db{\Delta}

\def\dbf{\Delta^{\rm f}}

\def\ethq{\varepsilon_q^{\rm th}}
\def\tth{t^{\rm th}}

\begin{document}
\title{Dynamic metastability in the two-dimensional Potts ferromagnet}
\author{Miguel Ib\'a\~nez Berganza$^{1}$\footnote{miguel.berganza@roma1.infn.it}, Alberto Petri$^{2}$, Pietro Coletti$^3$}
\affiliation{$^1$IPCF-CNR, UOS Roma {\em Kerberos} {\rm and} Dipartimento di Fisica, Universit\`a ``La Sapienza''.  Piazzale A. Moro, 5, 00185 Roma, Italy.\\ $^2$ Istituto dei Sistemi Complessi - CNR, via del Fosso del Cavaliere 100, 00133 Roma, Italy. \\ $^3$Dipartimento di Matematica e Fisica, Universit\`a Roma Tre, Via della Vasca Navale 84, 00146 Roma, Italy.}

\begin{abstract}

We investigate the non-equilibrium dynamics of the 2D Potts model on the square lattice after a quench below the discontinuous transition point. 
By means of numerical simulations of systems with $q =12,24$ and $48$ we observe the onset of a stationary regime below the temperature-driven transition, in a temperature interval decreasing with the system size and increasing with $q$. These results obtained dynamically agree with those obtained from the analytical continuation of the free energy \cite{Meunier2000Condensation}, from which metastability in the 2D Potts model results to be a finite size effect.

\end{abstract}

\maketitle

\section{Introduction}
When a liquid is cooled fast enough below its melting temperature, crystallization can be avoided, and the liquid enters a phase called supercooled \cite{debenedetti2001supercooled,debenedetti1996metastable,jerome1997dynamics,Jackle1986Models}. The supercooled phase is \textit{metastable}, it has a finite lifetime and it is unstable with respect to large fluctuations although those characteristics may not represent a practical limitation: Metastable states are ubiquitous \cite{debenedetti2001supercooled}, and not distinct from stable states in many practical respects. Metastability, as a general concept, is present in many fields of physics, from superconductivity to high-energy physics (see references in \cite{Rikvold1995Recent}). In particular, the understanding of metastability is crucial in the context of the glass-transition \cite{kauzmann1948nature,debenedetti2001supercooled,Jackle1986Models}: The structural glass transition occurs, during the cooling process, at a certain temperature below which the liquid falls out-of-equilibrium and forms the structural glass. However, and despite its ubiquity and intrinsic theoretical interest, metastability is still not well understood, and a lot of theoretical effort has been dedicated to the problem in the last decades \cite{Rikvold1995Recent,Binder1987Theory}. \\
From a purely thermodynamic point of view, the metastable phase is described by a free energy $f_{\rm m}(T,h)$ (where $T$ is the temperature and $h$ is the pressure, in the case of a fluid, or the magnetic field, in the case of a magnet), in a region of $(T,h)$ in which it coexist with the stable solution of the equation of state, $f<f_{\rm m}$. Metastable states satisfy the local stability condition, $(\partial_{hh} f_{\rm m})_T\le 0$, which is necessary, but not sufficient for stable equilibrium: Differently with respect to the stable state free energy $f$, metastable states are not stable with respect to a large enough fluctuation. Statistical mechanics in the mean field approximation, as the Landau theory of magnetism or the van der Waals equation of state for the gas-liquid condensation, accounts for metastable phases, which, in this context, exhibit the thermodynamic properties described above. In mean field approximation, the only allowed form of fluctuation is a spatially uniform change of the order parameter, and the free energy cost of such a change is extensive. For this reason, the free energy barriers separating the relative minima of the free energy from the absolute minima, the stable phase, are infinite in the thermodynamic limit, and, hence, the corresponding metastable state would have an infinite lifetime.\\
When fluctuations are taken into account, however, statistical mechanics of short-range interaction systems cannot account properly for metastability, since, when the thermodynamic limit is taken, the partition function is dominated by the global minimum of the free energy functional in phase space. Beyond mean field, there always exists a finite probability of surmounting the free energy barriers in phase space by a local nucleation process and, hence, a statistical description is only possible by the use of {\it restricted ensembles}, from which nucleated configurations are excluded \cite{Capocaccia1974,Penrose1971,Stillinger1995,Corti1995Metastability,Langer1967Theory,Langer1968Theory}. \\
An alternative is the dynamic description of metastability \cite{Binder1987Theory,Binder1973Scaling}, which is characterized by a two-step relaxation during the dynamic evolution of the system in the coexistence region. The two-step relaxation is associated with two times ($\tau_R \ll \tau$), such that the order parameter and other observables are stationary in time in the interval $\tau_R<t<\tau$, the lifetime of the metastable phase, $\tau$, being related in some way to the probability of nucleating the stable phase. For spin systems, nucleation times can be computed by the numerical solution of a master equation describing their temporal evolution, as pioneered by Binder for the Ising model \cite{Binder1973Scaling,Binder1974Investigation,Rikvold1995Recent}.
Although this dynamic definition of metastability is different from the reduced ensemble techniques mentioned before, there exist connections between the ensemble and the dynamical approaches \cite{Binder1987Theory,Rikvold1995Recent,Penrose1971,Gunther1993Numerical,Gunther1994Application}. Of particular relevance is the result by Langer \cite{Langer1968Theory}, who shows that, for a wide class of models, whose dynamics can be described by a Fokker-Plank equation, the nucleation rate, $\Gamma$, the number of nucleation events per unit time and volume, can be written under certain conditions as $\Gamma\sim \beta \kappa|{\rm Im}\, \tilde f|$, where $\beta$ is the inverse temperature, the prefactor $\kappa$ containing all dependence on the specific dynamics, and $\im \tilde f$ is the imaginary part of the analytical continuation of the equilibrium free energy $f$ in the unstable region of $(T,h)$. 
Analytically continuing the free energy beyond the transition point is equivalent to restricting the ensemble to under-critical droplets, in terms of Fisher's theory \cite{Fisher1967Theory}. Such an analytical continuation of the free energy has been computed for the first-order transition of the field-driven Ising model (or the equivalent lattice-gas model) by Langer \cite{Langer1967Theory,Gunther1980Goldstone}, based on Fisher's droplet approximation, and on a field-theoretical description of the free energy. The mentioned references provide an expression for $\Gamma$ as a function of (small) field and temperature in general dimensionality. This result has been extensively checked against Monte-Carlo local dynamics \cite{Binder1974Investigation,Rikvold1994Metastable,Binder1973Scaling,Heerman1984Nucleation,Binder2013Monte,Novotny2000Simulations,novotny2002large,Kolesik2003extreme}.  From these works, a good agreement between the theory and the numerics emerges in two, three and higher dimensions. \\
Things are much less clear in the case of the $q$-color Potts Model (PM) temperature-driven transition, which is discontinuous for $q>4$. Since Binder presented the problem in 1981 \cite{Binder1981Static}, it has been faced several times. Numerical results suggest the existence of a dynamic metastable phase for $q=5$, $d=2$ and $q=3$, $d=3$ \cite{Arkin1999,Arkin2000}, even if the metastable phase is not analyzed quantitatively (see also \cite{Berg2004,Ferrero2005,Velytsky2003}). In two dimensions, hysteresis cycles are studied numerically  \cite{Gupta1994}, and it is concluded that relaxation towards the equilibrium state occurs via nucleation. On the other hand, pseudo-critical attempts, finding evidences of second-order divergences at under-transition temperatures, suggest the existence of a nonzero spinodal limit \cite{Fernandez1992,Shulke2000}. This picture is confirmed by short-time approaches \cite{Loscar2009}, and by a recent study on large lattices \cite{Ferrero2012}, in which the authors report numerical evidence of the finiteness of the (disordered) energy slope at the transition temperature. The disappearance of the metastable interval for large sizes emerges instead in a Langer-like approach: an analytical continuation of the free energy within the droplet theory was done in 2000  \cite{Meunier2000Condensation} for the 2D $q$-PM. For finite-size systems, there is an under-transition temperature range where a convex Energy Probability Density (EPD) is found. Such a temperature interval, associated with a metastable state, is shown, however, to shrink to zero in the large system size limit \cite{Meunier2000Condensation}. This behavior is not present in the Ising model/lattice gas case, in which the metastable endpoint and the lifetime of the metastable phase become size-independent for sufficiently large sizes. In particular, this happens when the linear size $L$ is much larger than the length-scales involved in the nucleation processes:  the critical nucleating radius and the typical distance between critical clusters \cite{Rikvold1994Metastable}. Differently with respect to the Ising case, there is no {\it microscopic} droplet theory for the Potts case (the droplet expansion \cite{Meunier2000Condensation} is done in terms of macroscopic quantities), and it is missing a microscopic explanation of the disappearance of the metastable interval for large sizes. In any case, the equivalent nucleation mechanism would be size-influenced, or non-local, and hence essentially different in the Potts case, this difference possibly being present in other temperature-driven transitions.\\
This anomalous size-dependent behavior of the metastable states in the PM has been recently faced in \cite{Nogawa2011Static}. In this paper, the Monte Carlo (MC) dynamics of the 2D PM is studied by means of the typical passage time of the order parameter below a threshold, and it is signaled the existence of a finite-size inverse temperature $\beta_{\rm s}(N)>\bc$ (where $\bc$ is the transition temperature), separating different dynamical regimes. 
This anomalous behavior further motivates a dynamical study of metastability in the PM, allowing for a dynamical comparison with the droplet theory in \cite{Meunier2000Condensation}. The interest of such an approach has been pointed out in references \cite{Meunier2000Condensation,Nogawa2011Static}, since it could help clarifying to what extent the shrinking of the metastable interval is also observed in the dynamical scheme.\\ 
In the present work we show that the finite size effects described in \cite{Meunier2000Condensation} are indeed observable during the MC local dynamics. To this aim, we have developed a method to estimate the metastable endpoint, based on first-passage energy times. The so-obtained stationary temperature endpoint is shown to behave qualitatively as the pseudo-spinodal point of \cite{Meunier2000Condensation}, as a function of $q$ and $N$. \\
In the next section we review the main results of the droplet approach. Section \ref{sec:results} is to present our method and results. We conclude in section \ref{sec:conclusions}.


\section{Model and review of the Droplet Theory}
\label{sec:droplet}

The Potts model is one of the better known models in statistical physics \cite{Wu1982Potts}. It is defined on a lattice given by the adjacency matrix  $\cal A$ in which every site, $i=1,\ldots, N$, can take $q$ equivalent values, $\sigma_i=1,\ldots, q$, usually called \textit{colors}. The Hamiltonian is:

\begin{equation}
H=\frac{1}{2}\sum_{i,j}(1-\delta_{\sigma_i,\sigma_j})\,{\cal A}_{i,j}
\label{Hamiltonian}
\end{equation}
When ${\cal A}$ corresponds to an infinite square lattice with nearest-neighbor interactions, the model is known to present a first-order phase transition for $q>4$, and a continuous phase transition for $q\le4$, separating a paramagnetic high-temperature phase, in which all colors coexist in equal proportion, from a ferromagnetic $q$-degenerated low-temperature phase. On the square lattice both transitions occur at a critical inverse temperature ${\bc}(q)=\ln(1+q^{1/2})$. Due to the presence of many competing ground states, the Potts model can exhibit non-equilibrium features well different from the Ising model \cite{Ferrero2005,Petri2008,Loureiro2012,Olejarz2013}.\\
Requiring agreement with exact results on the first-order transition in two dimensions, the authors of \cite{Meunier2000Condensation} postulate the form of the free energy of the disordered phase, expressed as a Fisher sum ($\db=\beta-\bc<0$) \cite{Meunier2000Condensation}:

\begin{equation}
F(\db) \propto \sum_{a=1}^\infty a^{-\tau} e^{\db a-\omega a^\sigma}
\label{eq:Fdroplet}
\end{equation}
The exponents are fixed to $\tau=7/3$, $\sigma=2/3$, by matching previous results for the correlation length at the transition, and the known value of the $\alpha$, $\nu$ critical exponents for $q=4$. Each term in Eq. (\ref{eq:Fdroplet}) is proportional to the partition function of the ensemble of clusters of area $a$. The continuum limit to Eq. (\ref{eq:Fdroplet}) is a function, $\phi$, analytic for ${\db}<0$, i.e., in the disordered phase region. For $\db>0$, $\phi$  diverges, but its analytic continuation to complex inverse temperatures can be evaluated for positive $\db$. The successive step is to obtain the finite-size EPD at the transition point, $P_{N,0}(\varepsilon)$ ($\varepsilon$ being the energy per site) by a Lapace transform of $\phi$ (which requires integrating  $\phi$ on a complex contour). Finally, the EPD in the metastable interval $P_{N,\db}(\varepsilon)$, with $\db>0$ is further obtained  by re-weighting: $P_{N,\db}(\varepsilon)=P_{N,0}(\varepsilon)\,e^{-\db N \varepsilon}$ \cite{Meunier2000Condensation}. The EPD $P_{N,\db}$ is such that, for $N$ and $\db>0$ fixed, there exist a local minimum $\varepsilon_{\rm m}(\db,N)<\varepsilon^{(d)}$, such that $P_{\db,N}$ represent stable states for $\varepsilon>\varepsilon_{\rm m}$. The bound $\varepsilon_{\rm m}$ implies a finiteness of the associated metastable state lifetime. The position of the minimum increases with $\db$, in such a way that there is an inverse temperature $\bc+\db^*(N)$ above which the EPD is no longer convex. Due to an anomalous size dependence of $P$ for energies lower than the equilibrium disordered energy, $\varepsilon^{(d)}$, the value $\db^*(N)$ shrinks to zero for large $N$, with the law $\sim N^{-1/3}$, as can be calculated approximately \cite{Berganza2014}. 
The theory predicts, in this way, that the convex-EPD describing metastable states for $\db>0$, is not but a finite-size effect. \\
The free energy $\phi$ and the EPD $P_{\db,A}$ in \cite{Meunier2000Condensation} are formulated in terms of scaling energy, temperature and area variables, independent of $q$, which are related to physical variables by products of powers of the correlation length at the transition point $\xi_q$, in which the whole $q$-dependence is enclosed. The result is such that the metastable interval increases with $q$. In particular, for fixed $N$ and sufficiently low $\db$, the metastable energy interval endpoint behaves as $\varepsilon_{\rm m}\sim -(3\db\, \xi_q/2)^{-3}$, as can be derived from \cite{Meunier2000Condensation} in saddle-point approximation, valid for $\db\searrow 0$.  Since $\xi_q$ is a decreasing function, the interval increases with $q$,  and the inverse temperature endpoint (let us call its explicit $q$-dependence $\db^*_q(N)$) consequently increases with $q$.    In conclusion, according to the theory, $\db_q^*(N)$ increases with $q$ and decreases with $N$.

\section{Dynamical method and results}
\label{sec:results}

\begin{figure}[t!]           
\begin{center} 
 \includegraphics[width=.9\columnwidth]{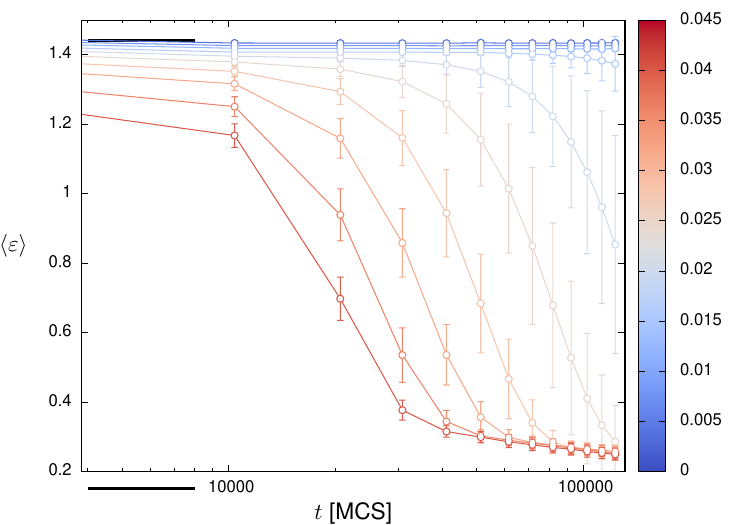} 
\caption{Average energy per site $\<\varepsilon\>$ (over different dynamical sequences) versus time of the $q=24$ PM after a quench to different quench depths $\db$ (in colorbar). The errorbars indicate the standard deviation over different realizations of the dynamics. The horizontal segments indicate the exact equilibrium energies of both phases at $\bc$, $\varepsilon^{(d)}$ and $\varepsilon^{(o)}$.}
\label{fig:et}
\end{center}   
\end{figure}                          

\begin{figure}[t!]           
\begin{center} 
 \includegraphics[width=.8\columnwidth]{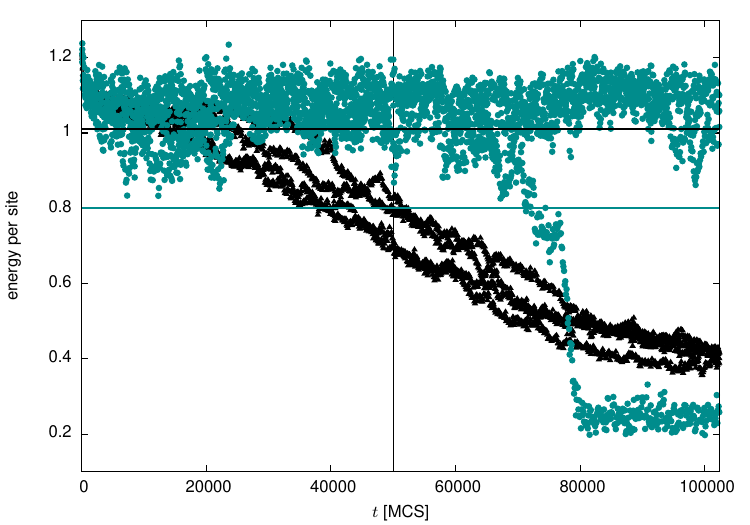} 
 \includegraphics[width=.8\columnwidth]{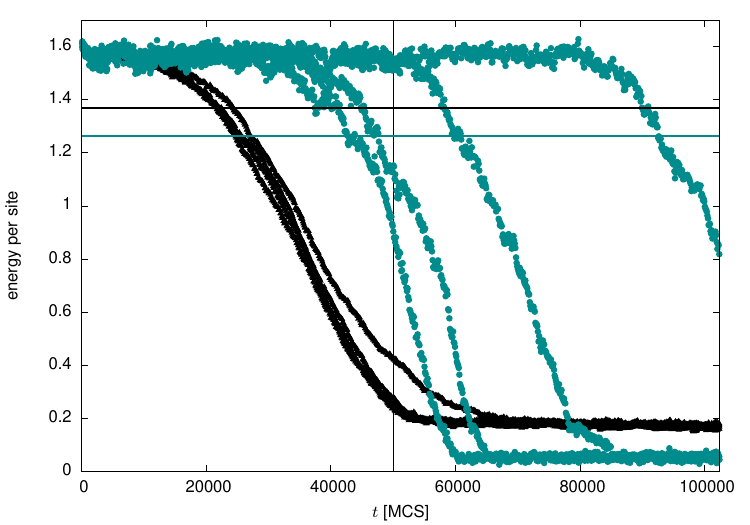} 
\caption{Upper panel: four random realizations of the dynamics for $q=12$, $\db=0.008$, $L=64$ (green points) and $L=256$ (black points). The horizontal lines mark the thresholds $\ethq(N)$ (lower for lower $N$) and the vertical line indicates $\tth$ (see text). Lower panel: idem with $q=48$, $\db=0.063$.}
\label{fig:reals}
\end{center}   
\end{figure}

In Fig. \ref{fig:et} we show the behavior of the energy per site  $\varepsilon$ versus time in a system with $q = 24$, $L = 128$, for several values of the quench inverse temperature interval, $\db$. A completely uncorrelated configuration is used as initial condition. By using different realizations of the initial condition and of the random sequence used in the Metropolis algorithm we have computed the first two moments, $\<\varepsilon\>$ and $\<\varepsilon^2\>- \<\varepsilon\>^2$, of the non-equilibrium EPD for $\db>0$, where the average is over different instancies of the MC dynamics. For small values of $\db$ we see that $\<\varepsilon\>$ stays about constant and close to the high energy disordered phase, $\varepsilon^{(d)}$, up to $t > 10^5$ MCS, indicating a long average lifetime of the metastable state. For larger $\db$, the two-step relaxation characterizing dynamic metastability is observed. The metastable lifetime decreases for increasing $\db$, up to become smaller that the simulation time, and $\<\varepsilon\>$ is seen to relax towards a value close to the low energy ordered phase, $\varepsilon^{(o)}$. At the same time, the second moment shows to be small when $\<\varepsilon\>$ is close to $\varepsilon^{(d)}$ or $\varepsilon^{(o)}$, and large during the relaxation. This is an indication of the fact that different realizations with the same $\db$ can follow very different energy trajectories while relaxing towards the ordered state, and therefore display a different lifetime. An example of this fact is shown in the bottom panel of Fig. \ref{fig:reals}, where three different realizations of a system with $L=256$ and $q=48$ are quenched at $\Delta=0.0063$, and are seen to decay at different times. In systems with short-range interactions, the lifetime of the metastable phase is a stocastic quantity and, as a consequence, the averages shown in Fig. 1 could be not the more suitable quantity for determining a possible size scaling of the metastable phase. \\
A strong finite-size effect, as the one described in the precedent section, is indeed immediately observed also in single realizations of the dynamical evolution, as illustrated in Fig. \ref{fig:reals}, where we show the energy per site of single instancies of the MC dynamics as a function of the number of Monte Carlo steps (MCS) with a local (Metropolis) algorithm, at fixed values of the quench depth $\db>0$. Sufficiently small systems present an energy plateau, while larger systems do not: for them the shown $\db$ is presumably larger than the metastable endpoint $\db^*(N)$. A possible way to estimate numerically the length of the metastable interval is to compute {\it the fraction of realizations presenting an energy plateau}, among a set of many MC sequences of configurations, generated from different initial conditions and sequences of random numbers. The metastable interval can be arbitrarily defined in this way as the $[\bc:\bc+\dbf]$ interval in which the fraction stays above a given threshold. We note that such an arbitrariness is unavoidable in finite-dimensional systems where, at variance with respect to mean-field systems, the dynamical endpoint of metastability is a stochastic, and time-dependent concept, and can be defined only in average even for a fixed lifetime. \\
Of course the largest arbitrariness is in the definition of plateau. Here we choose a criterion for stationarity, based on two arbitrary quantities: we target a realization as stationary whenever its energy per site remains larger than a threshold $\ethq(N)$, for a time longer than a time threshold $\tth$. A more sophisticated criterion, which allows for a more accurate comparison with the theory, will be presented elsewhere \cite{Berganza2014}.\\
Let us describe the details of our method. We have studied the $q$-PM with $q=12,24,84$, on square lattices of length $L=64,128,256,384$, $N=L^2$, with periodic boundary conditions. For each $q$ and $L$, we perform series of Metropolis MC sequences of configurations, starting from different random configurations and differing also in the random number sequence. We have performed a minimum of 200 realizations and a maximum of 800, depending on $q$ and $L$, up to $1.024\,10^5$ MCS. The energy per site is computed each 128 MCS. This is done for ten values of $\db_n=\bc+n\, \delta\beta$, where $n=1,\ldots,10$ and $\delta\beta=0.001$, $0.004$ and $0.007$ for $q=12,24,48$, respectively. Afterwards, we count how many realizations stay above $\ethq(N)$ for at least $\tth$ MCS. In this way, for example, none of the $L=256$ realizations of Fig. \ref{fig:reals}, upper panel, are considered as stationary, while 3/4 of the $L=64$ do. In this way we estimate the {\it fraction of realizations presenting a plateau}, $\rho_q(\db,N)$. A key point in our approach is that the threshold $\ethq(N)$ differs from the stationary (time-averaged) energy by a quantity which decreases with $N$, so that we take into account fluctuations (decreasing with size) around the stationary value. Fixing the threshold energy independently of $N$ would lead instead to take as stationary realizations whose time-averaged energy is actually decreasing for large systems or, on the other hand, to target as non-stationary small size realizations that are indeed stationary but present large fluctuations. We therefore set    $\ethq(N)=\varepsilon^{(d)}_q-c/N^{1/2}$, being $c$ a constant: The threshold energy differs from the equilibrium energy at $\bc$ by a quantity proportional to the fluctuations of the energy $[\<\varepsilon^2\>-\<\varepsilon\>^2]^{1/2} \sim N^{1/2}$. Fixing the reference disordered energy as $\varepsilon_q^{(d)}$ we are implicitly assuming that the energy averaged over stationary realizations at $\db>0$ (expected to coincide with the average of the EPD $P_{\db,N}$, performed for $\varepsilon>\varepsilon_{\rm m}$) does not differ too much with respect to the average transition energy at $\db=0$, $\varepsilon^{(d)}_q$. We have verified that our results are essentially independent with respect to such small variations on the value of the reference energy, and hence the only arbitrary constant  is $\tth$. Moreover, choosing a sufficiently small $\tth$,  the resulting $\rho_q$ functions are quite robust against $\tth$. The discriminated plateaus in our scheme, whose lifetime is lower than $\tth$, are essentially not influential.\\
\begin{figure}[t!]           
\begin{center} 
 \includegraphics[width=.9\columnwidth]{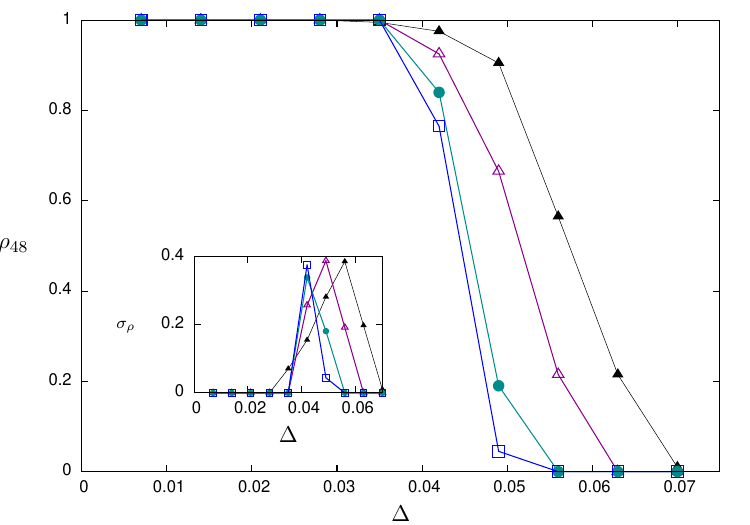} 
\caption{Fraction of realizations presenting a plateau, $\rho(\db,N)$, after a quench at inverse temperature $\bc+\db$, for $q=48$ and $N=64^2$, $128^2$, $256^2$ and $384^2$ (from right to left). The inset shows the standard deviation $\sigma_\rho$ of the distribution of fractions.}
\label{fig:fracs}
\end{center}   
\end{figure}                          
Our results for $\rho(\db,N)$ are illustrated in Fig. \ref{fig:fracs} for $q=24$, the other cases are qualitatively identical. The behavior is as follows: for low values of $\db$, immediately above the inverse transition temperature, all realizations present a metastable energy plateau. For large values of $\db$, well above the  metastable endpoint $\bc+\db^*$, no realization presents a plateau, and the crossover occur at smaller $\db$'s for larger system sizes. Fixing $\rho$ at an arbitrary value, one gets an estimate of the temperature endpoint of the metastable phase: $\dbf_q(N)$. In Fig. \ref{fig:temps} we present the obtained values, for a fixed value of $\rho=0.5$. Although, due to the arbitrariness described above,  $\dbf_q(N)$ is not an accurate estimate of $\db_q^*(N)$ of the Meunier and Morel theory \cite{Meunier2000Condensation}, its qualitative behavior turns out to be the same, thus providing a dynamical confirmation of the theory: $\dbf_q(N)$ increases with $q$ and monotonically decreases with $N$. Moreover, a detailed in progress analysis \cite{Berganza2014} both of the $\db^*(N)$ obtained from the theoretical EPD, and of its corresponding numerical dynamical estimation (averaging over properly stationary sequences only) is yielding values of $\db_{12}^*(N)$ rather similar to the $\dbf_{12}(N)$ that we have obtained here. However, due to the existence of realizations that are stationary despite being beyond the validity limit of the theory $\db^*$, of which we have provisional numerical evidence \cite{Berganza2014}, $\dbf$ could be an over-estimation of $\db^*$ for large sizes, although both quantities present the same qualitative behavior.

\begin{figure}[t!]           
\begin{center} 
 \includegraphics[width=.9\columnwidth]{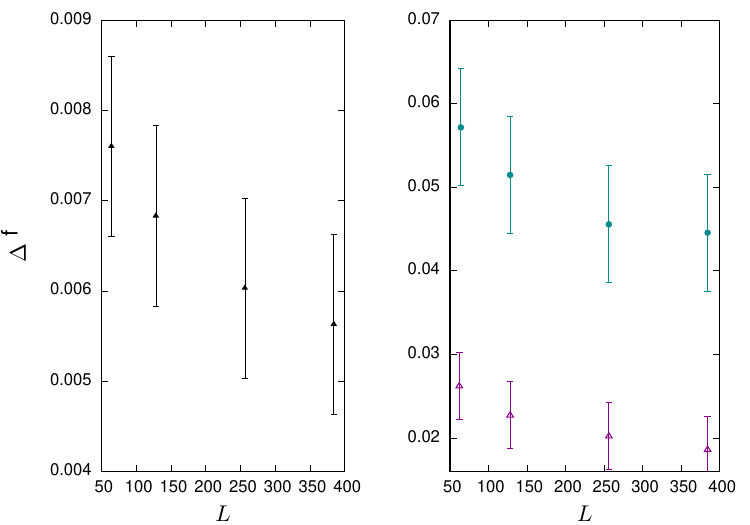} 
\caption{Left: Estimated values of $\dbf_{q}$ for $q=12$. Right: idem, for $q=24$ and $48$ (from bottom to top).}
\label{fig:temps}
\end{center}   
\end{figure}


\section{Conclusions}
\label{sec:conclusions}

The existence of metastable states in the Potts model is even now a debated problem in statistical physics since three decades. In this article we have shown how in a dynamical MC scheme the interval in which stationarity is present shrinks with the system size $L$, in qualitative agreement with the theoretical work \cite{Meunier2000Condensation}. The method we have used is general and may serve for the study of other models. Despite its simplicity, it reproduces reasonably well the endpoint $\db^*_q$ predicted by the theory.\\
The fast vanishing of the metastable regime presented here can be understood as a call for caution when interpreting hysteresis cycles performed for the $q$-PM: if the step, $\delta \beta$, of the cooling at a rate $\delta \beta / \Delta t$ is larger than the corresponding size-dependent metastable interval, the points of the hysteresis diagram would not correspond to metastable states, but rather to heterogeneous, non-equilibrium configurations, which have already nucleated.\\
The  finite-size effect illustrated in this article challenges for the search of a kind of microscopic, but size-dependent nucleation-like mechanism, that would be essentially different with respect to the well understood lattice gas case, and that may constitute a different paradigm, perhaps  also present in other first-order transitions. The theory \cite{Meunier2000Condensation} is based on the behavior of the finite-size EPD, which is obtained by an inverse Laplace transform of the infinite-volume free energy $\phi$, so that it does not provide an evident microscopic interpretation in terms of droplets. In the Ising model nucleation, the bulk term in the free energy of the ensemble of clusters of fixed area is given by the external field, a size-independent quantity, while in the PM it may come from an entropy-maximizing constraint which keeps the permutation symmetry of colors unbroken. In  systems with increasing size, a single cluster would be less confined by such an effective field since it contributes less to the global magnetization of its corresponding color. We propose to perform a droplet calculation, starting from a size-dependent expression for the free energy, Eq. (\ref{eq:Fdroplet}), in terms of size-dependent quantities, possibly confirming the  microscopic mechanism proposed above, and reproducing the phenomenology predicted by \cite{Meunier2000Condensation}\\

\section{Acknowledgments} M. I. acknowledges discussions with  Andrea Cavagna,  Alessandro Attanasi and Fabrizio Antenucci.\\

\bibliography{berganza-potts}
\bibliographystyle{apsrev4-1}

\end{document}